\documentclass[cits,proceedings]{PoS}
\pdfoutput=1
\usepackage{amsmath}
\usepackage{mathbbol}
\usepackage{placeins}
\usepackage{braket}

\usepackage{feynmp}
\ifpdf
\fi

\title{Static-light meson-meson potentials}
\ShortTitle{Static-light meson-meson potentials}

\author{Gunnar Bali, \speaker{Martin Hetzenegger}  \\
		Institut f\"ur Theoretische Physik, Universit\"at Regensburg,\\ 
		93040 Regensburg, Germany\\
		E-mail:
		\email{gunnar.bali@physik.uni-regensburg.de},
		\email{martin.hetzenegger@physik.uni-regensburg.de}\\\vspace*{.25cm}}

\author{
\rm\centering (QCDSF Collaboration)}

\abstract{We investigate potentials between pairs of static-light mesons
in $N_{\mathrm f}=2$ Lattice QCD, in different spin channels.
The question of attraction and repulsion is particularly interesting with
respect to the $X(3872)$ charmonium state and charged candidates such as
the $Z^{+}(4430)$. We employ the nonperturbatively improved
Sheikholeslami-Wohlert fermion and the Wilson gauge actions at
a lattice spacing $a\approx 0.084$ fm and
a pseudoscalar mass $m_{\mathrm PS}\approx 760$ MeV.
We use stochastic all-to-all propagator techniques, improved by
a hopping parameter expansion. 
The analysis is based on the variational method, utilizing
various source and sink interpolators.}

\FullConference{The XXVIII International Symposium on Lattice Field Theory\\
		June 14-19, 2010\\
		Villasimius, Sardinia, Italy}
\begin{document}
\section{Introduction}
We numerically determine ground and excited states of
static-light mesons (${\mathcal B}=Q\bar{q}$)  as well as
intermeson potentials between pairs of static-light 
mesons, ${\mathcal B}({\mathbf r}){\mathcal B}({\mathbf 0})$ and
${\mathcal B}({\mathbf r})\overline{\mathcal B}({\mathbf 0})$, with a
static quark-quark (or quark-antiquark)
separation $r=|{\mathbf r}|=Ra$. $a$ denotes the lattice spacing,
$Q$ a static colour source and
the positions of the light quarks $q$ are not fixed.
For large heavy quark masses the spectra of 
heavy-light mesons are determined by
excitations of the light quark and gluonic degrees of
freedom. In particular, the vector-pseudoscalar
splitting vanishes and the static-light meson
${\mathcal B}$ can be interpreted as either a
$\overline{B}$, a $\overline{B}^*$, a $D$ or a $D^*$
heavy-light meson.

Static-light intermeson potentials were first evaluated on the lattice
by Michael and Pennanen in the quenched approximation~\cite{Michael:1999nq}
and with $N_{\mathrm f}=2$ Sheikholeslami-Wohlert
sea quarks~\cite{Pennanen:1999xi}.
A more detailed
quenched study can be found in ref.~\cite{Detmold:2007wk} and
the un-quenched case is revisited with twisted mass fermions
by Wagner in these
proceedings~\cite{Wagner:2010ad}. Meson-antimeson potentials
were computed in ref.~\cite{Pennanen:1999xi} and, with
Wilson sea quarks, in ref.~\cite{Sesam}.
Graphically, the quark line diagrams that we evaluate can be depicted as,
\begin{align}
\parbox{.03\textwidth}{
\begin{fmffile}{diag1}
\begin{fmfgraph*}(10,40)
\fmfstraight
\fmfkeep{static-light}
\fmfleft{lu,lo}
\fmfright{ru,ro}
\fmf{fermion}{lu,lo}
\fmf{wiggly,tension=0,left=0.5}{lo,lu}
\fmf{phantom_arrow,tension=0,left=0.5}{lo,lu}
\end{fmfgraph*}
\end{fmffile}}
\;\;,\;\;\qquad
\parbox{.08\textwidth}{
\begin{fmffile}{diag2}
\begin{fmfgraph*}(30,40)
\fmfstraight
\fmfkeep{static-light}
\fmfleft{lu,lo}
\fmfright{ru,ro}
\fmfbottom{lu,um,ru}
\fmf{fermion}{lu,lo}
\fmf{fermion}{ru,ro}
     \fmflabel{$\leftarrow$ $R$ $\rightarrow$}{um}
\fmf{wiggly,tension=0,left=0.5}{lo,lu}
\fmf{phantom_arrow,tension=0,left=0.5}{lo,lu}
\fmf{wiggly,tension=0,right=0.5}{ro,ru}
\fmf{phantom_arrow,tension=0,right=0.5}{ro,ru}
\end{fmfgraph*}
\end{fmffile}}
\;\;,\;\;\;\quad
\parbox{.08\textwidth}{
\begin{fmffile}{diag3}
\begin{fmfgraph*}(30,40)
\fmfstraight
\fmfkeep{static-light}
\fmfleft{lu,lo}
\fmfright{ru,ro}
\fmfbottom{lu,um,ru}
\fmf{fermion}{lu,lo}
\fmf{fermion}{ro,ru}
     \fmflabel{$\leftarrow$ $R$ $\rightarrow$}{um}
\fmf{wiggly,tension=0,left=0.5}{lo,lu}
\fmf{phantom_arrow,tension=0,left=0.5}{lo,lu}
\fmf{wiggly,tension=0,left=0.5}{ru,ro}
\fmf{phantom_arrow,tension=0,left=0.5}{ru,ro}
\end{fmfgraph*}
\end{fmffile}}\;\; ,
\label{equ:feynman1} \\ \nonumber
\end{align}
where the static-light correlation function is shown on the left,
and those for the meson-meson and meson-antimeson
cases in the 
centre and on the right, respectively.
Straight lines represent static quark propagators and
wiggly lines light quark propagators. In this paper we neglect
explicit meson exchange (i.e.\ ``box'' and ``cross'') diagrams.
The analyses of these as well as of a larger lattice
volume are in progress. This means that here we only
consider the isospin $I=0$ ${\mathcal B}{\mathcal B}$
and the $I=1$ ${\mathcal B}\overline{\mathcal B}$ combinations.

The static-light correlation function in Euclidean time
$t=Ta$ is given by,
\begin{align}
C(t) = \langle  0 | 
	(\overline{Q}_{\alpha} \; \mathcal{O}_{\alpha \beta} \; 
q_{\beta})_{x+t\hat{4}} \;
	(\bar{q}_{\gamma} \; \mathcal{O}_{\gamma\delta} \;
Q_{\delta})_{x} 
	| 0 \rangle 
	= \left\langle  \; \mathrm{Tr} \; \left[ 
	 \prod^{T -1}_{k=0} \,
U^{\dagger}_{x +  ka\hat{4},4} \; \frac{\Eins + \gamma_4}{2} \;\mathcal{O} \;
	M^{-1}_{x+t\hat{4},x} \; \mathcal{O}
	\right] \right\rangle_{\!\!\!U} \; .
	\label{equ:correlator}
\end{align}
The trace is over colour (not displayed) and Dirac indices,
$\left\langle \cdot\right\rangle_U$ indicates 
the expectation value over gauge configurations and
$\hat{\mu}$ denotes a unit vector in $\mu$-direction.
We average over all possible source points
$x/a \in \{1,\ldots,L_{\sigma}\}^3\times
\{1,\ldots,L_{\tau}\}$ to reduce statistical errors
where $x=({\mathbf x},x_4)$.
The correlator is automatically zero-momentum
projected since
${\mathbf x}$ is the same at source and sink, due to the
static propagator.
$M^{-1}_{yx}=\langle q_y\bar{q}_x\rangle$ is the
propagator for the light quark $q$ on a given gauge configuration
and $U_{x,\mu}$ is the gauge link connecting the lattice
site $x$ with $x+a\hat{\mu}$. The absence of the
spin in the static propagator necessitates the
$(\Eins+\gamma_4)/2$ Dirac projection of the (fermionic)
static-light ``meson'' to fix the parity $P$. This is very
similar to baryonic correlation functions where a spin $\frac12$
source is created by three (rather than one) light quarks.
Meson-(anti)meson correlation functions can be obtained by
combining the above correlator with another one that is
spatially shifted by a distance ${\mathbf r}$, before taking
the gauge average.

\section{Representations and classification of states}
In the continuum limit, the static-light states can be
classified according to fermionic representations $J^P$
of the rotation group ${\mathrm O(3)}$.
At vanishing distance ${\mathbf r}={\mathbf 0}$ the
${\mathcal B}{\mathcal B}$ and ${\mathcal B}\overline{\mathcal B}$
states can be characterized by
integer $J^P$ and $J^{PC}$ quantum numbers, respectively.
However at $r=|{\mathbf r}|>0$ the $\mathrm{O(3)}$
(or $\mathrm{O(3)}\otimes{\mathcal C}$) symmetry is broken
down to its cylindrical $\mathrm{D_{\infty h}}$ subgroup.
The irreducible representations of this
are conventionally labelled by the spin along the axis
$\Lambda$, where $\Sigma,\Pi,\Delta$ refer to $\Lambda=0,1,2$,
respectively, with a subscript $\eta=g$ for gerade (even)
$PC=+$ or $\eta=u$
for ungerade (odd) $PC=-$ transformation properties with respect
to the midpoint.  All
$\Lambda\geq 1$ representations are two-dimensional. The
one-dimensional $\Sigma$ representations carry an
additional $\sigma_v=\pm$ superscript
for their reflection symmetry
with respect to a plane that includes the two endpoints.
 
To create states of different $J^{P(C)}$ we use operators 
$\mathcal{O}$ that contain combinations of Dirac
$\gamma$-matrices and covariant lattice
derivatives $\nabla[U]$ that act on a fermion spinor
$q$ as,
\begin{align}
\nabla_{\mu} q_x = U_{x,\mu} q_{y+a\hat{\mu}} -
U_{x,-\mu} q_{x-a\hat{\mu}}\,, \quad 
	\mbox{where} \quad U_{x,-\mu} = U^{\dagger}_{y - a\hat{\mu},\mu}\,.
\end{align}
On the lattice the continuum rotational
symmetry is broken and the groups
$\mathrm{O(3)}$ and $\mathrm{D_{\infty h}}$ need to be replaced
by their finite dimensional
subgroups $\mathrm{O_h}$ and $\mathrm{D_{4h}}$, respectively.
We label fermionic representations of the octahedral group
$\mathrm{O_h}$ as $\mathrm{O_h}'$. For fermionic
representations of $\mathrm{D_{\infty h}}$ that we do not
need in the present context, see ref.~\cite{Najjar:2009da}.
It is well known, see e.g.
ref.~\cite{Bali:2000gf}, that
the assignment of a continuum spin to a lattice result can be
ambiguous, in particular for radial excitations because a given
$\mathrm{O_h}$ representation can be subduced from several continuum
$J$s. For instance,
\begin{equation}
G_1 \leftarrow J = \frac{1}{2}, \frac{7}{2}, \dots\,,\quad
H   \leftarrow J = \frac{3}{2}, \frac{5}{2}, \dots\,,\quad
A_1 \leftarrow J = 0, 4,\dots\,,\quad
T_1 \leftarrow J = 1, 3, 4,\dots\,.
\label{equ:mapping} 
\end{equation}
For $\Lambda\leq 2$ the mapping of continuum
$\mathrm{D_{\infty h}}$ onto discrete $\mathrm{D_{4h}}$ representations 
is more straight forward. Hence in this case we
adopt the continuum notation only.

The operators that we used to create the static-light
mesons are displayed in table
\ref{tab:operators_sl} (see, e.g., ref.~\cite{Michael:1998sg}).
\TABULAR{|c|c|c|c|c|}{
\hline
	\textbf{$\mathcal{O}$} &wave~\protect\cite{Michael:1998sg}&
$\mathrm{O_h}'$ rep. &continuum $J^{P}$ & $J^{P}$ (heavy-light)\\
\hline
$\gamma_5$&$S$&$G_1^{+}$&$\frac{1}{2}^{+}$&$ 0^{-}, 1^{-}$\\
$\Eins$&$P_-$&$G_1^{-}$&$\frac{1}{2}^{-}$&$ 0^{+}, 1^{+}$\\
$\gamma_i\nabla_i$&$P_-$&$G_1^{-}$&$\frac{1}{2}^{-}$&$0^{+}, 1^{+}$\\
$\left( \gamma_1\nabla_1 - \gamma_2\nabla_2 \right) + \mathrm{cycl.}$&$P_+$& $H^{-}$   & $\frac{3}{2}^{-}$& $ 1^{+}, 2^{+}$  \\\hline}{Operators
and representations for static-light mesons. In the last
column we display the $J^P$ for a heavy-light meson, obtained by
substituting the (spinless) static source by a heavy fermion.\label{tab:operators_sl}}
\TABULAR{|c|l|l|l|l|}{\hline
$\mathcal{O} \; \otimes\; \mathcal{O}$&\multicolumn{2}{l|}
{${\mathcal B}{\mathcal B}$}&\multicolumn{2}{l|}
{${\mathcal B}\overline{\mathcal B}$}\\
	\cline{2-5}
 & $r=0$: $J^{P}$ & $r>0$: $\Lambda_\eta^{(\sigma_\nu)}$ &
$r=0$: $J^{PC}$ &$r>0$: $\Lambda_\eta^{(\sigma_\nu)}$ \\
	\hline
	$\gamma_5 \;\otimes\; \gamma_5$  &  $0^{+}, 1^+$  &  $\Sigma_g^{+},
\Pi_g$ &  $0^{++}, 1^{+-}$  &  $\Sigma_g^{+},\Sigma_u^-,\Pi_u$ \\
	$\Eins \;\otimes\; \Eins$  &  $0^{+}, 1^+$  &  $\Sigma_g^{+},
\Pi_g$ &  $0^{++}, 1^{+-}$  &  $\Sigma_g^{+},\Sigma_u^-,\Pi_u$ \\
	$\gamma_5 \;\otimes\; \Eins$     &  $0^{-}, 1^-$  &  $\Sigma_u^{-},\Pi_u$ &  $0^{-+}, 1^{--}$  &  $\Sigma_u^{-},\Sigma_g^+,\Pi_g$ \\
	$\gamma_5 \;\otimes\; \gamma_i\nabla_i$     &  $0^{-}, 1^-$  &  $\Sigma_u^{-},\Pi_u$ &  $0^{-+}, 1^{--}$  &  $\Sigma_u^{-},\Sigma_g^+,\Pi_g$ \\
	$\gamma_5 \;\otimes\; [\left( \gamma_1\nabla_1 - \gamma_2\nabla_2 \right) \; + \mathrm{cycl.}]$  &  $1^{-},2^-$  &  $\Sigma_u^{+}, \Pi_u,\Delta_u$ & $1^{--}, 2^{-+}$  &  $\Sigma_g^{+}, \Pi_g, \Sigma_u^+,\Pi_u,\Delta_u$\\
	$\gamma_i\nabla_i \;\otimes\; [\left( \gamma_1\nabla_1 - \gamma_2\nabla_2 \right) \; + \mathrm{cycl.}]$  &  $1^{+},2^+$  &  $\Sigma_g^{-}, \Pi_g,\Delta_g$ &  $1^{+-}, 2^{++}$  &  $\Sigma_u^{-}, \Pi_u, \Sigma_g^+,\Pi_g,\Delta_g$\\
	\hline
}{Operators and continuum representations for
the meson-meson (${\mathcal B}{\mathcal B}$) and
meson-antimeson (${\mathcal B}\overline{\mathcal B}$) potentials.
\label{tab:operators_potential}}
The intermeson potentials were obtained
by combining two static-light mesons of different (or the same)
quantum numbers. This can be projected into an irreducible $\mathrm{D_{\infty h}}$
representation, either by coupling the light quarks together in spinor
space~\cite{Wagner:2010ad} or by projecting the static-light meson
spins into the direction $\hat{\mathbf r}$
of the static source distance, by applying
$\frac12(\Eins\pm i\gamma_5\pmb{\gamma}\cdot\hat{\mathbf{r}})$,
and taking appropriate symmetric ($\Lambda_s=1$) or antisymmetric
($\Lambda_s=0$) spin combinations. These two approaches can be related
to each other via a Fierz transformation.
For the preliminary results presented here we have not yet performed
this projection and different representations will mix.
The analyzed operators and the corresponding representations are listed 
in table \ref{tab:operators_potential}.

For $J>0$ and $r>0$ the irreducible representations of
$\mathrm{O(3)}$ split up into two or more 
irreducible representations of $\mathrm{D_{\infty h}}$.
For instance the angular momentum of the $P_+$ operator within
our $1^-,2^-$ $r=0$ state can be perpendicular or parallel
to the intermeson axis. 
For the axis pointing into the $\hat{3}$-direction,
we call the $S\otimes P_+$ operator
$\gamma_5\otimes(\gamma_1\nabla_1 - \gamma_2\nabla_2)$ ``parallel"
($\|$) and the other combinations 
``perpendicular" ($\bot$). The 
$\bot$ state has no angular momentum pointing into the 
direction of the axis and
hence only couples with the light quarks to
$\Sigma_u^+$ and $\Pi_u$. Vice versa, the
$\|$ operator can only create $\Pi_u$ and $\Delta_u$ states
but not 
the $\Sigma_u^+$.

\section{Simulation and analysis}
\TABULAR{|c|l|l|l|l|l|l|l|}{
	\hline
	volume $L_{\sigma}^3\times L_{\tau}$   & $\beta$ & $\kappa_{\mathrm{val}} = \kappa_{\mathrm{sea}}$  & $c_{\mathrm{SW}}$ & $a/\mathrm{fm}$ & $La/\mathrm{fm}$ & $m_{\mathrm{PS}}/\mathrm{MeV}]$ & $N_{\mathrm{conf}}$ \\
	\hline
	$16^3\times32$ & $5.29$  & $0.13550$ & $1.9192$ & $0.084$  & $1.34$  & $760(3)$ & $200$ \\
	\hline}
{Lattice parameters.
\label{tab:lattice}}
We employ $N_{\mathrm f}=2$ Sheikholeslami-Wohlert configurations
generated by the QCDSF Collaboration~\cite{AliKhan:2003br}.
The parameter values are listed in table~\ref{tab:lattice}, where
the scale is set using $r_0(\beta,\kappa)=0.467$~fm.
The pseudoscalar mass corresponds to its infinite
volume value.
We use the Chroma software system~\cite{Edwards:2004sx}.

To achieve high statistics in the evaluation of the
diagrams eq.~(\ref{equ:feynman1}), all-to-all propagators
need to be computed.
This is done using stochastic estimator techniques,
see ref.~\cite{Bali:2009hu} and references therein.
We generate 300 complex ${\mathbb Z}_2$ noise sources and
apply the hopping parameter expansion to reduce the
stochastic variance~\cite{Thron,Sesam,Bali:2009hu}.
Furthermore we enhance the signal over noise ratio by
employing a static action with reduced self-energy~\cite{Sesam}.
This is done by applying one stout smearing step~\cite{Stout} with the
parameter $\rho=1/6$ to the temporal links, used to calculate
the static propagators.
Wuppertal smearing~\cite{Wuppertalsmear} is applied to the
source and sink operators, where we employ spatially
smeared parallel transporters~\cite{Sesam} with
the parameters $n_{\mathrm{iter}}=15$, $\alpha=2.5$.
The Wuppertal smearing hopping parameter
value $\kappa_{\mathrm w}=0.3$ is combined with three iteration
numbers $N_{\mathrm{iter}}\in\{16,50,100\}$.
Masses are then extracted from the resulting correlation matrices,
by means of the variational method~\cite{Vari}, solving a generalized 
eigenvalue problem. Errors are calculated using
the jackknife method.

\section{Results}
The eigenvalues $\lambda^{(k)}(t,t_0)$ of the generalized
eigenvalue problem~\cite{Vari},
\[
C_{ij}(t)u_j^{(k)}=\lambda^{(k)}(t,t_0)C_{ij}(t_0)u_j^{(k)}\,,\]
are fitted to one- and two-exponential ans\"atze, to obtain the $k$th
mass. The appropriate values of $t_0$ and the fit ranges in $t$
are determined from monitoring the effective masses,
\[
m_{\mathrm{eff},t_0}^{(k)}(t+a/2)=a^{-1}\ln\left(\frac{\lambda^{(k)}(t,t_0)}
{\lambda^{(k)}(t+a,t_0)}\right)\,.\]
On the left hand side of figure~\ref{fig:mp_para},
we display the effective ground state
energy levels $E_{\mathrm{eff}}$ for $t_0=2a$ of the 
$\gamma_5 \otimes \gamma_5$ ${\mathcal B}{\mathcal B}$
system for different distances $R=r/a$. In the limit $r\rightarrow\infty$
these will approach the sum of two $\frac{1}{2}^+$ static-light meson
effective masses (dotted curve). At short distances we see
attraction in this channel. The quality of the effective
mass plateaus deteriorates with decreasing distance
since the wavefunction of
such interacting states becomes
more than the mere product of our two static-light meson interpolators.
We should also keep in mind that so far we did not perform the
singlet spin projection and hence there will be additional pollution
from the $\Pi_g$ state, see table~\ref{tab:operators_potential}. 
For the example displayed, we perform two-exponential fits
to the $t/a\in\{4,\ldots,10\}$ data to obtain the masses.

\FIGURE
{\includegraphics[width=.47\textwidth,clip]{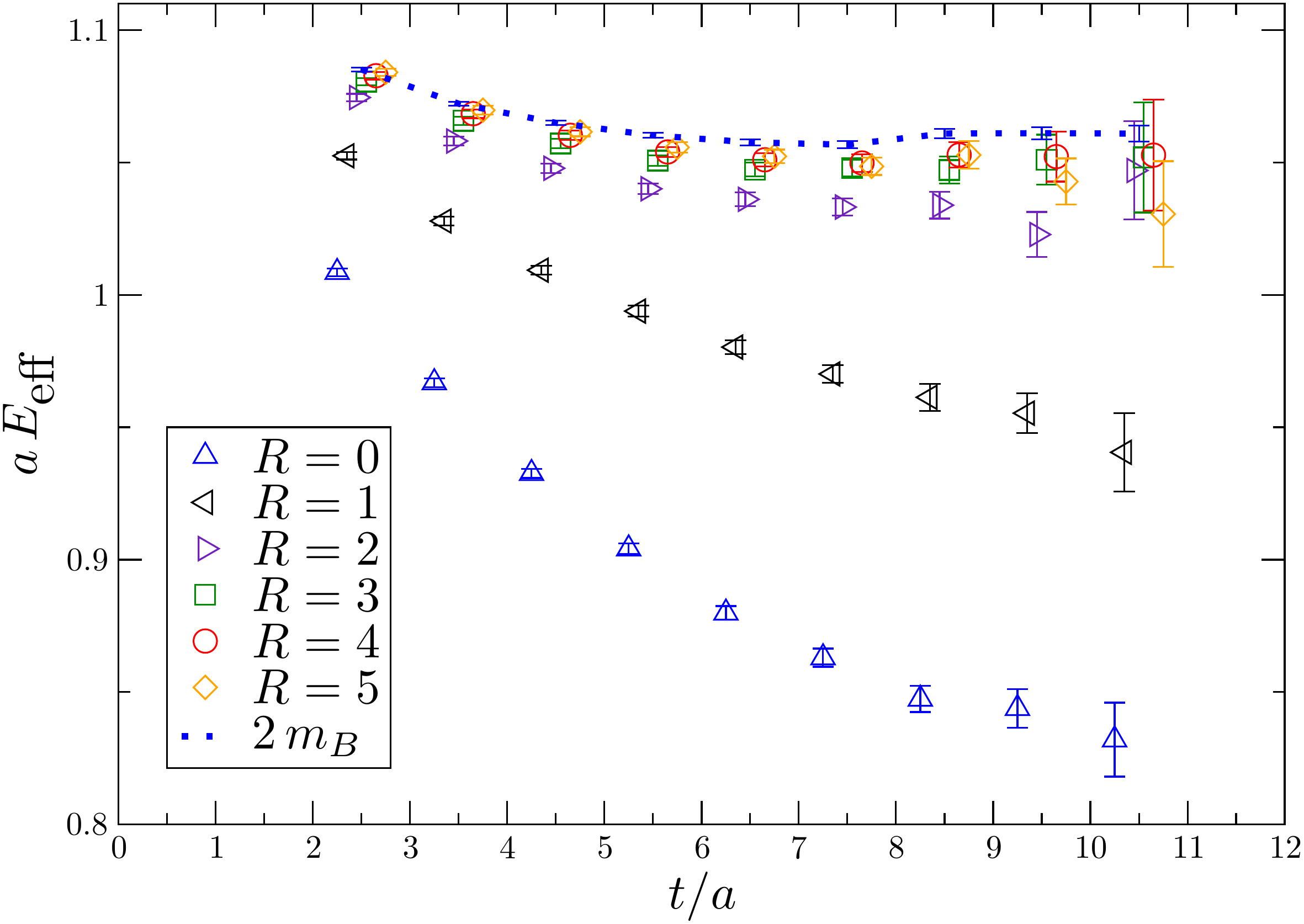}\hspace*{0.03\textwidth}
	\includegraphics[width=.47\textwidth,clip]{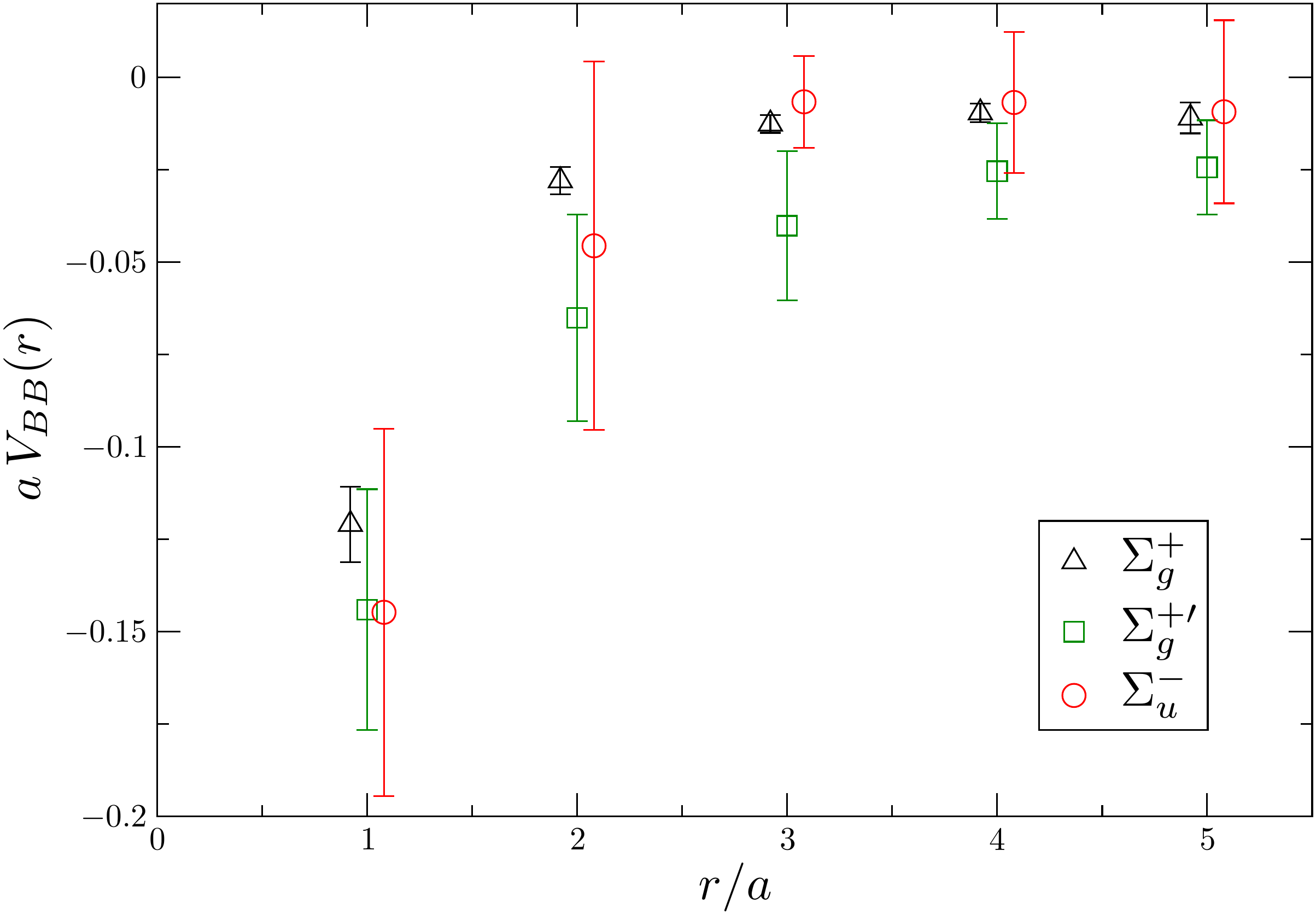}
\caption{Effective groundstate masses of the 
$\gamma_5 \otimes \gamma_5$ operator at different distances
(left hand side).
The dotted line corresponds to twice this mass for a single
$\frac12^+$ static-light meson.
The intermeson potential $V(r)$ for the combinations
$\gamma_5 \otimes \gamma_5$ ($\Sigma_g^+, \Sigma_g^{+\prime}$)
and $\gamma_5 \otimes 1$ ($\Sigma_u^-$) (right hand side).
\label{fig:mp_para}}}
We define intermeson potentials as the difference between
the meson-meson energy levels and the
$r\rightarrow\infty$ two static-light meson limiting cases:
\begin{equation}
V_{\!\mathcal{B}_1\mathcal{B}_2}(r)=E_{\!\mathcal{B}_1\mathcal{B}_2}(r)
-\left(m_{\!\mathcal{B}_1}+
m_{\!\mathcal{B}_2}\right)\quad\stackrel{r\rightarrow\infty}{\longrightarrow}\quad 0\,.
\end{equation}
The results for the groundstate ($\Sigma_g^+$) and the first
excited state ($\Sigma_g^{+\prime}$) of the 
$\gamma_5 \otimes \gamma_5$ operator
and the $\Sigma_u^-$ groundstate of $\gamma_5 \otimes \Eins$
are plotted on the right hand side of figure \ref{fig:mp_para}.
In all these $I=0$ channels there is attraction of the order of
50~MeV at a distance of $0.2$~fm.
 
\FIGURE
{\includegraphics[width=.7\textwidth]{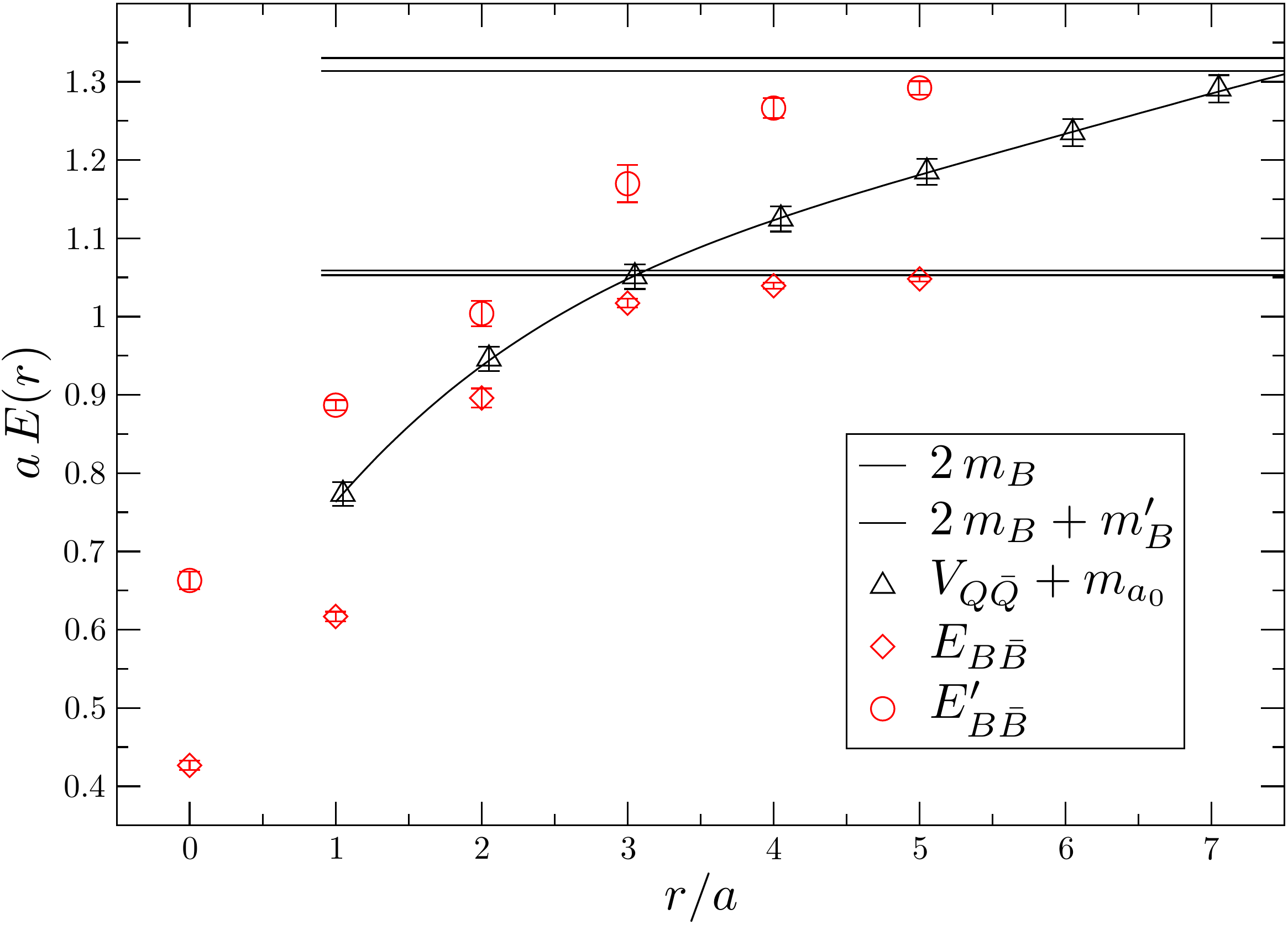}
\caption{Effective masses for
the $\gamma_5 \otimes \gamma_5$ $\mathcal{B}\overline{\mathcal B}$
ground and first excited states (red symbols). The horizontal lines
denote sums of two static-light meson masses while the black curve
and symbols are the static $Q\overline{Q}$ potential, offset by
the scalar $a_0$ meson mass.
\label{fig:mp_split_antipara}}}

In figure \ref{fig:mp_split_antipara} we display the ground state
and first excited state energy levels for
the ${\mathcal B}\overline{\mathcal B}$ meson-antimeson case
in the $\gamma_5 \otimes \gamma_5$ channel. The two
horizontal lines correspond to twice the ground state mass of the
$\frac12^+$ static-light meson and to the sum of its ground and its
first
excited state masses, the expected $r\rightarrow\infty$ limits.
At first sight there appear to be very substantial short distance
attractive forces in this channel.
However, states consisting of a $Q\overline{Q}$ static potential
and a scalar $I=1$ particle will have the same quantum numbers.
For our lattice parameters the $a_0$ meson is the lowest such
state, with masses of two pseudoscalars as well as of a $P$-wave
vector lying higher.
We include the sum
of these two masses in the figure. The ground state
${\mathcal B}\overline{\mathcal B}$
energy still lies below this level but to decide
whether we effectively see the sum of $a_0$ and the static potential
and to disentangle which $I=1$ $\Sigma_g^+$ energy level is the
lowest one, interactions of
the $a_0$ with the static potential will have to be taken into account.
We hope that simulations on a larger volume will help
to clarify this.

\section{Conclusions}
We investigated interactions between pairs of static-light mesons
and found attraction in the $I=S=0$ sector. Meson-antimeson
potentials are also very interesting with respect to
charmonium threshold states~\cite{Brambilla:2010cs}
($D\overline{D}$ molecules or tetraquarks)
but difficult to disentangle from
mesons that are bound to static-static states
(hadro-quarkonium~\cite{Dubynskiy:2008mq}).
We are in the process to extend our study to $I=1$ ${\mathcal B}{\mathcal B}$
and to $I=0$ ${\mathcal B}\overline{\mathcal B}$ states and to disentangle
$S_{\hat{\mathbf r}}=\pm 1$ from $S_{\hat{\mathbf r}}=0$ states.
Also simulations on a larger volume are in progress.

\acknowledgments
We thank Sara Collins, Christian Ehmann, Christian Hagen and Johannes Najjar
for their help. We also thank Dirk Pleiter and other members of the
QCDSF Collaboration for generating the gauge ensemble.
The computations were mainly performed on Regensburg's Athene HPC cluster.
We thank Michael Hartung and other support staff. We acknowledge
support from the
GSI Hochschulprogramm (RSCHAE), the Deutsche Forschungsgemeinschaft
(Sonderforschungsbereich/Transregio 55) and the
European Union (grants 238353, ITN STRONGnet
and 227431, HadronPhysics2).

\end{document}